\documentclass[amsmath,amssymb,aps,prc,twocolumn,floatfix,nofootinbib,superscriptaddress,notitlepage]{revtex4-2}
\usepackage{graphicx}% Include figure files
\usepackage{dcolumn}% Align table columns on decimal point
\usepackage{bm}% bold math
\usepackage{multirow,booktabs}
\usepackage[table]{xcolor}
\usepackage{enumitem}
\usepackage{mathtools}
\usepackage{caption}
\usepackage{calc}
\usepackage{cancel}
\usepackage{framed}
\usepackage{subcaption}
\usepackage{float}
\usepackage{ragged2e}
\usepackage{soul}

\usepackage[normalem]{ulem}  % \sout{old text} for strikeout
\usepackage{color} % For blue in-text comments and additions
\renewcommand{\sout}{\bgroup \color{red} \ULdepth=-.5ex \ULset}
\newcommand{\rot}{{\rm{rot}}}

\newcommand{\OPE}{{\text{OPE}}}
\newcommand{\vev}[1]{\langle 0|{{#1}}|0\rangle}

\begin{document}
\title{Spin structure  of spin-1 charmonium states near $T_c$}
\author{HyungJoo Kim}
%\email{hugokm0322@gmail.com}
\affiliation{International Institute for Sustainability with Knotted Chiral Meta Matter (WPI-SKCM${}^2$), Hiroshima University, Higashi-Hiroshima, Hiroshima 739-8526, Japan}

\date{\today}

\begin{abstract}
We investigate the spin structure of the \(1^{--}\) and \(1^{++}\) charmonium states near the critical temperature using QCD sum rules. To this end, we compute the contribution of the dimension-4 twist-2 gluon operator to the two-point function of heavy vector and axial vector currents in a rotating frame. As temperature increases, the quark spin contribution slightly increases, while the quark orbital angular momentum decreases by a comparable amount. The gluon contribution remains nearly unchanged. These thermal changes cancel each other, ensuring that the total spin is preserved even at finite temperature.
\end{abstract}

\maketitle
\section{Introduction}
Charmonium, a bound state of a charm quark and its antiquark, plays a crucial role in probing the properties of strongly interacting matter under extreme conditions. At high temperatures, hadronic matter undergoes a phase transition to a deconfined state known as the quark-gluon plasma (QGP). During this process, charmonium states undergo spectral modifications and may eventually dissolve at sufficiently high temperatures. Ever since Matsui and Satz proposed that the suppression of charmonium production in heavy-ion collisions could be a signal of QGP formation~\cite{Matsui:1986dk}, the thermal behavior of charmonium has attracted significant attention. In particular, its spectral modifications have been extensively studied using various theoretical approaches, including lattice QCD simulations and effective field theories~\cite{Umeda:2002vr,Asakawa:2003re,Digal:2001ue,Brambilla:2008cx,Rothkopf:2011db}.
For comprehensive reviews, see Refs.~\cite{Brambilla:2010cs,Brambilla:2014jmp,Rothkopf:2019ipj}.

Historically, the QCD sum rule (QCDSR) has also played an important role in the study of charmonium spectral properties. This method was developed shortly after the discovery of the \( J/\psi \) and successfully predicted the mass of the \( \eta_c \), which was later confirmed by experiments~\cite{Shifman:1978bx,Shifman:1978by,Shifman:1978zq}. Since then, QCDSR studies of charmonium states have been further refined to improve its predictive power~\cite{Reinders:1981si,Bertlmann:1981he,Reinders:1984sr} and extended to explore thermal modifications, including mass shifts and broadening near the critical temperature~\cite{Morita:2007hv,Morita:2009qk,Gubler:2011ua,Gubler:2018ctz}. More recently, we have applied this approach to study hadron spin decomposition \cite{Kim:2022vtt,Kim:2025iyq}. In Ref.~\cite{Kim:2022vtt}, we analyzed the spin structure of spin-1 charmonium states in vacuum and found that the quark spin contributes about 90\% in the \( J^{PC} = 1^{--} \) state and about 40\% in the \( 1^{++} \) state. A comparable trend has also been reported in a recent lattice QCD simulation of charmed hadrons~\cite{He:2024irz}.

As with other spectral properties, the charmonium spin structure is also expected to undergo modifications at finite temperature. Motivated by this, we investigate how the spin structure of the vector \((1^{--})\) and axial vector \((1^{++})\) charmonium states changes near the critical temperature in this work. 
Our analysis essentially follows the methodology developed in Refs.~\cite{Kim:2022vtt,Kim:2025iyq}. Here, we extend the previous work to finite temperature by newly incorporating the dimension-4 twist-2 gluon operator into the operator product expansion.  
As temperature increases, we observe the common behavior for both vector and axial vector channels: the quark spin contribution increases by a few percent, whereas the quark orbital angular momentum decreases by a comparable amount. The two effects largely cancel out each other, while the gluon contribution remains nearly unchanged.

The paper is organized as follows. In Sec.~\ref{sec:OPE}, we present the Wilson coefficients for the dimension-4 twist-2 gluon operator.  
Sec.~\ref{sec:QCDSR} outlines the method for extracting the spin components of the ground-state charmonium based on the QCDSR approach.  
Sec.~\ref{sec:analysis} is devoted to numerical analysis. 
The paper is summarized and concluded in Sec.~\ref{sec:summary}.  
In the Appendix, we display the Wilson coefficients of other operators for completeness, along with the numerical values of some parameters used in this work.

\section{Operator Product Expansion}\label{sec:OPE}
To study the spin structure of spin-1 charmonium states, we consider the two-point function of the heavy vector or axial vector  current in a rotating frame.
As demonstrated in Refs.~\cite{Kim:2022vtt,Kim:2025iyq}, the explicit spin decomposition of the two-point function can be obtained by  the following expression:
\begin{align}
    \hspace*{-0.02cm}
    \Pi^{\mu\nu}(q)=i\int d^4x d^4y e^{iqx}\vev{T\{j^\mu(x)M_z(y)j^\nu(0)\}},
    \label{methoda}
\end{align}
where $M_z$, the total angular momentum density of quarks and gluons in $z$-direction, is inserted into the typical two-point function defined in an inertial frame. Throughout this work, we adopt a convention in which $M_z$ is decomposed into three gauge invariant components \cite{Ji:1996ek}: 
\begin{align}
\hspace*{-0.14cm}
\vec{M}(x)=\frac{1}{2} \bar{\psi} \vec{\gamma} \gamma^5 \psi + \psi^\dagger (\vec{x} \times (-i\vec{D})) \psi+ \vec{x} \times (\vec{E} \times \vec{B}), \label{jqcd}
\end{align}
where the first and second terms correspond to quark's spin ($S_q$) and orbital angular momentum ($L_q$), respectively, while the third term corresponds to gluon total angular momentum ($J_g$). Here, \(\psi\) denotes the quark field, \(\vec{D} = \vec{\partial} + ig\vec{A}\) the covariant derivative, and \(\vec{E}\) and \(\vec{B}\) the color electric and magnetic fields, respectively, with Dirac and color indices omitted.

In the deep Euclidean region, where \( Q^2 = -q^2 \gg 0 \), we perform an operator product expansion (OPE)  of Eq.~\eqref{methoda} at leading order in $\alpha_s$ and up to dimension-4 gluon operators: scalar $G_0$ and twist-2 $G_2$ which are defined by
\begin{align}
    \langle 0| \frac{\alpha_s}{\pi} G^{\mu\alpha} G^\nu{}_{\alpha} |0\rangle =g^{\mu\nu}G_0+  ( u^\mu u^\nu - \frac{1}{4} g^{\mu\nu} ) G_2,
\end{align}
where $u^\mu$ represents the four-velocity of a medium, implying that the twist-2 component exists only in a medium.
By collecting all contributions, we confirm that the OPE results satisfy the following expression:
\begin{align}
    \hspace*{-0.02cm}
    \Pi^{\mu\nu}(q) &= \left[ i(g^{1\mu}g^{2\rho} - g^{1\rho}g^{2\mu}) + (\vec{q} \times i\vec{\partial}_q)_z\, g^{\mu\rho} \right] \nonumber\\
    &\quad \times \partial_q^0 \left\{ i \int d^4x\, e^{iqx} \left\langle T\{ j_\rho(x)\, j^\nu(0) \} \right\rangle \right\},
    \label{methodb}
\end{align}
where the first and second terms inside the brackets \([\,\cdots\,]\) correspond to the total spin and the total orbital angular momentum of the system, respectively, while the expression inside the braces \(\{\,\cdots\,\}\) represents the two-point function  in an inertial frame.
Therefore, the equivalence between Eq.~\eqref{methoda} and Eq.~\eqref{methodb} demonstrates that the total angular momentum is conserved.  
Through this equivalence, we gain insight into how the quark and gluon angular momenta contribute to the total spin and orbital angular momentum of the system as described by the two-point function.  
Furthermore, for the first time, we demonstrate that the OPE of a non-scalar operator also satisfies this equivalence, even in the presence of a medium.

To focus on the spin part, we isolate the right-circularly polarized component in the rest frame and define
\begin{align}
    \Pi^\rot(\omega^2) = \frac{i\,\Pi^{12}(\omega)}{\omega},
\end{align}
where \( \omega \) denotes the total energy of the system. In this work, the medium is also assumed to be at rest, i.e., \( u^\mu = (1,0,0,0) \). The OPE series can then be written as
\begin{align}
    \hspace*{-0.08cm}
    \Pi^\rot_\text{OPE}=\Pi^{\text{pert}}(Q^2)+C^{4,0}(Q^2)\cdot G_0+C^{4,2}(Q^2)\cdot G_2,\label{OPE0}
\end{align}
where \( \Pi^{\text{pert}} \) and \( C^{4,0} \) are the Wilson coefficients for the identity operator and the dimension-4 scalar gluon condensate \( G_0 \), respectively, which were computed in our previous work~\cite{Kim:2022vtt}.  
The coefficient \( C^{4,2} \) corresponds to the dimension-4 twist-2 gluon condensate \( G_2 \), and is newly computed in this work.
The detailed spin decomposition of \( C^{4,2} \) is given as follows:
\hfill\break
\hfill\break
\noindent $\bullet$ Vector channel: 
\begin{align}
    C_{S_q}^{4,2}&=\frac{1}{12Q^4}\big\{2+y+12J_1-72J_2+94J_3-36J_4\big\},\\
    C_{L_q}^{4,2}&=\frac{-1}{12Q^4}\big\{17-y-22J_1-2J_2+6J_3\big\},\\
    C_{J_g}^{4,2}&=\frac{-1}{12Q^4}\big\{11+2y-(6+4y)J_1-9J_2+4J_3\big\},
\end{align}
\noindent $\bullet$ Axial vector channel: 
\begin{align}
    C_{S_q}^{4,2}&=\frac{1}{12Q^4}\big\{6+y-36J_1+48J_2-18J_3\big\},\\
    C_{L_q}^{4,2}&=\frac{-1}{12Q^4}\big\{25-y-30J_1-9J_2+14J_3\big\},\\
    C_{J_g}^{4,2}&=\frac{-1}{12Q^4}\big\{7+2y+(2-4y)J_1-13J_2+4J_3\big\},
\end{align}
where $y=Q^2/m^2$, $J_n(y)=\int^1_0 dx [1+x(1-x)y]^{-n}$, and \( m \) denotes the heavy quark mass. This completes the spin decomposition of the OPE series up to dimension-4 and leading order in \(\alpha_s\).

%As a comparison, the OPE of the vacuum invariant function
%\(\Pi_\vac(q) = i \int d^4 x\, e^{iqx} \VEV{T\{ j^\mu(x)\, j_\mu(0) \} }\)
%can be written as
%\begin{align}
%    \hspace*{-0.2cm}
%    \Pi_\vac^\OPE = \Pi^\text{pert}_\vac(Q^2) + C^{4,0}_\vac(Q^2) \cdot G_0 + C^{4,2}_\vac(Q^2) %\cdot G_2,
%\end{align}
%where the Wilson coefficients are well known and can be found in the literature~\cite{...}.  
%As shown in Ref.~\cite{Kim:2022vtt}, this vacuum result is directly related to the OPE of the two-%point function in the rotating frame via
%%\begin{align}
 %   \Pi_\rot^\OPE(Q^2) = -2 \frac{\partial \Pi_\vac^\OPE(Q^2)}{\partial Q^2}. \label{equivalence}
%\end{align}

\section{QCD Sum Rules}\label{sec:QCDSR}
The spin structure of the ground-state charmonium can be studied using the spin-decomposed OPE series within QCDSR framework. As a first step, we extract the ground-state contribution from the two-point function as follows:
\begin{align}
   \bar{\Pi}^\rot(Q^2,s_0)= \Pi^\rot_\OPE(Q^2)-\frac{1}{\pi}\int^\infty_{s_0}ds\frac{\text{Im}\Pi^\text{pert}(s)}{s+Q^2},\label{pibar}
\end{align}
where the continuum contribution is approximated by the perturbative part above the threshold parameter \( s_0 \). To improve the convergence of the OPE series and further suppress the remaining continuum contribution, we apply the Borel transform to Eq.~\eqref{pibar}, which is defined as
\begin{align}
 \mathcal{B}\equiv\lim_{\substack{Q^2/n \rightarrow M^2, \\ n,Q^2
  \rightarrow \infty}}
  \frac{\pi(Q^2)^{n+1}}{n!}\left(-\frac{d}{dQ^2}\right)^n,
\end{align}
where \( M \) is the Borel mass parameter.  This yields the Borel-transformed sum rule:
\begin{align}
    \mathcal{M}(M^2,s_0)%&\equiv \mathcal{B}[\bar{\Pi}^\rot(Q^2,s_0)]\\
    &=\int^{s_0}_{4m^2}ds \,e^{-s/M^2}\text{Im}\Pi^\text{pert}(s)\nonumber\\
    &\quad+\widetilde{C}^{4,0}(M^2)\cdot G_0+\widetilde{C}^{4,2}(M^2)\cdot G_2,
   % &=\mathcal{M}_{S_q}+\mathcal{M}_{L_q}+\mathcal{M}_{J_g}
\end{align}
where $\mathcal{M}\equiv\mathcal{B}[\bar{\Pi}^\rot]$, and \( \widetilde{C}^{4,0}(M^2) \) and \( \widetilde{C}^{4,2}(M^2) \) are the Borel-transformed Wilson coefficients. From the spin-decomposed OPE, 
the above sum rule can be decomposed into three distinct parts, \( \mathcal{M}_{S_q} \), \( \mathcal{M}_{L_q} \), and \( \mathcal{M}_{J_g} \) according to their spin origins.
We then examine the individual contributions of \( S_q \), \( L_q \), and \( J_g \) relative to the total $\mathcal{M}$:
\begin{align}
    s_q(M^2,s_0)=\frac{\mathcal{M}_{S_q}(M^2,s_0)}{\mathcal{M}(M^2,s_0)},\\
    l_q(M^2,s_0)=\frac{\mathcal{M}_{L_q}(M^2,s_0)}{\mathcal{M}(M^2,s_0)},\\
    j_g(M^2,s_0)=\frac{\mathcal{M}_{J_g}(M^2,s_0)}{\mathcal{M}(M^2,s_0)},
\end{align}
where \( s_q \), \( l_q \), and \( j_g \) reveal the spin components of the ground-state charmonium at rest, corresponding to the contributions of the quark spin, quark orbital angular momentum, and total gluon angular momentum, respectively. 

Although the total sum of these components is always unity, the individual components depend on the Borel mass $M$ and the threshold parameter $s_0$.
In practical analyses, it is necessary to determine a reliable range of the Borel mass, known as the Borel window, along with an appropriate continuum threshold. 
The Borel window is typically defined by imposing two (somewhat empirical) conditions. The lower bound \( M_{\text{min}} \) is chosen such that the contributions from the \( \mathcal{O}(\alpha_s) \) perturbative correction and the dimension-4 operators do not exceed 30\% of the total OPE. This criterion ensures the convergence of the truncated OPE series. The upper bound \( M_{\text{max}} \) is set by requiring that the ground-state contribution exceeds 60\% of the total OPE, thereby ensuring sufficient suppression of the continuum. 

It should be noted, however, that the spin decomposition of the \( \mathcal{O}(\alpha_s) \) correction  
has not yet been performed in the present analysis.  
%Consequently, the Borel window cannot be reliably determined based solely on our spin-decomposed OPE.  
Therefore, following the strategy employed in Ref.~\cite{Kim:2022vtt},  
we simply adopt the Borel window determined from the OPE series of the two-point function 
defined in an inertial frame, which include  the \( \mathcal{O}(\alpha_s) \) corrections~\cite{Reinders:1984sr}.  
The continuum threshold is then determined by requiring that the charmonium mass extracted from this OPE  
be as stable as possible with respect to \( M \) within the Borel window. The resulting Borel window and threshold parameter are then consistently employed in our spin decomposition analysis and are listed in Appendix~\ref{appendix2}. For further details, see Refs.~\cite{Morita:2009qk,Gubler:2018ctz,Kim:2022vtt,Kim:2023ihl}.

\section{Numerical Analysis} \label{sec:analysis}
\begin{figure}[h]
    \centering
    \includegraphics[width=1\linewidth]{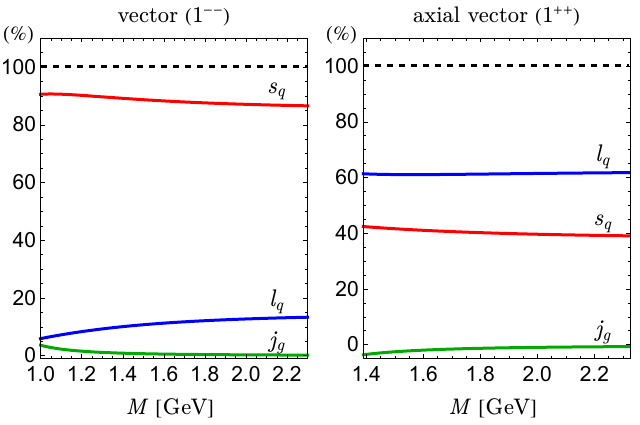}
    \caption{\justifying Spin components of the ground state in vacuum are plotted as functions of \( M \) for the vector channel (left) and axial vector channel (right). Dashed lines indicate the total spin sum, which is 100\% in all cases.}
    \label{fig:plotvac}
\end{figure}

We now investigate the spin structure of the vector \((1^{--})\) and axial vector \((1^{++})\) charmonium states.  
As a starting point, we revisit their spin structures in vacuum, which were previously studied in Ref.~\cite{Kim:2022vtt}.  
In contrast to the previous work, the quark orbital angular momentum is expressed here as a single gauge-invariant term.
For the input parameters, we use the charm quark mass  
\( m_c(p^2 = -m_c^2) = 1.262 \,\mathrm{GeV} \) and  
the strong coupling constant \( \alpha_s(8m_c^2) = 0.21 \) \cite{Morita:2009qk}.  
In vacuum, the scalar gluon condensate is taken as \( G_0 = (0.35\,\mathrm{GeV})^4 \),  
while \( G_2 =0\). 
In Fig.~\ref{fig:plotvac}, the individual spin components for the vector and axial vector channels  are plotted as functions of the Borel mass \( M \). The Borel window and  the threshold parameter are listed in Appendix~\ref{appendix2}.
To read off the spin component values, we take the average over the Borel window and present the results in Table~\ref{tab:vac}, with uncertainties given as standard deviations.
\begin{table}[h]
    \caption{Charmonium spin structure in vacuum}
    \centering
    \begin{tabular}{ccc}
    \toprule
    Component(\%) & Vector &Axial vector \\
    \midrule
    $s_q$ &\; \,87.6$\pm$1.5 &40.1$\pm$1 \\
    $l_q$ & 11.8$\pm$2& \;\, 61.4$\pm$0.3\\
    $j_g$ & \;\,\,\,\,\;0.6$\pm$0.5& \;\,\,$-$1.5$\pm$0.7\\
    \bottomrule
    \end{tabular}
    \label{tab:vac}
\end{table}

\begin{figure*}[!htbp]
    \centering
    \includegraphics[width=0.69\linewidth]{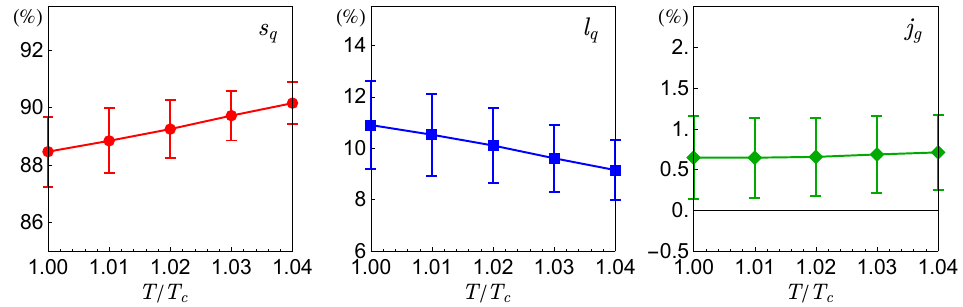}
    \caption{Thermal modification of the charmonium spin structure in the vector channel}
    \label{fig:vectorT}
\end{figure*}
\begin{figure*}[!htbp]
    \centering
    \includegraphics[width=0.69\linewidth]{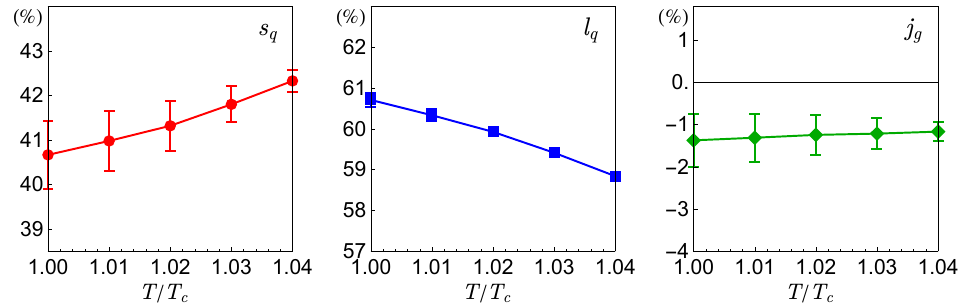}
    \caption{Thermal modification of the charmonium spin structure in the axial vector channel}
    \label{fig:axialT}
\end{figure*}

We now extend this analysis to finite temperature, incorporating the twist-2 gluon operator \( G_2 \). 
Our focus is on the region near the critical temperature $T_c$, specifically from \( 1.00\,T_c \) to \( 1.04\,T_c \). 
%Since the temperature scale is much lower than the typical separation scale of the OPE, which is of the order of \( \sim 1\,\mathrm{GeV} \), 
Here, we assume that all temperature dependence is encoded in the change of the gluon condensates,  
while the Wilson coefficients remain unaffected~\cite{Hatsuda:1992bv}. The numerical values of \( G_0 \) and \( G_2 \) in this temperature region are taken from pure SU(3) lattice gauge simulations~\cite{Boyd:1996bx, Morita:2009qk}.  
We then repeat the same analysis at different temperatures, using the Borel window and continuum threshold listed in Appendix~\ref{appendix2}.

The temperature dependence of the individual spin components  in the vector channel is shown in Fig.~\ref{fig:vectorT}. Error bars in the figure indicate standard deviations. As temperature increases, we observe that the quark spin contribution \( s_q \) increases by a few percent, while the quark orbital angular momentum contribution \( l_q \) decreases by a similar amount. 
The gluon total angular momentum \( j_g \) remains nearly unchanged. Here, the total variations cancel each other exactly, so the total spin of  charmonium state is preserved even at finite temperature. A similar trend is also observed in the axial vector channel, as shown in Fig.~\ref{fig:axialT}. Although the error bar appears very small for \( l_q \) in the axial vector channel, it should be noted that the present analysis does not account for uncertainties in other input parameters and condensate values.

%On the other hand, it would be interesting to extend the analysis to even higher temperatures.  In the non-relativistic quark model, the vector (\(1^{--}\)) state has a spin structure of \( (s_q : l_q : j_g) = (100 : 0 : 0) \), while the axial vector (\(1^{++}\)) state has \( (50 : 50 : 0) \).  At first glance, the spin structures of both states appear to approach these model predictions as the temperature increases.  However, QCDSR analyses with only dimension-4 operators are known to break down beyond the vicinity of the critical temperature~\cite{Morita:2007hv, Morita:2007pt, Morita:2009qk}.  Therefore, a reliable extension to higher temperatures would require the inclusion of higher-dimensional operators and a better understanding of their thermal behavior.

\section{Summary and conclusion} \label{sec:summary}
In this work, we newly compute the contribution of the dimension-4 twist-2 gluon operator to the two-point function of heavy vector and axial vector currents in a rotating frame. We also confirm that angular momentum is conserved even in the OPE involving non-scalar operators.
By incorporating this twist-2 gluon contribution into the previous work, we examine the spin structures of the vector ($1^{--}$) and axial vector ($1^{++}$) charmonium states near the critical temperature using the QCDSR approach.
Our analysis reveals that, as temperature increases, the quark spin contribution increases, while the quark orbital angular momentum decreases by a comparable magnitude. The gluon angular momentum contribution remains nearly unchanged. These variations cancel each other, ensuring that the total spin is preserved even at finite temperature.
This thermal behavior may be understood as a consequence of weakened binding between heavy quarks at higher temperatures. 
As the system gradually transitions from a strongly bound state toward two nearly free quarks, the orbital angular momentum originally stored in the bound state may be partially transferred to the spin  of the individual quarks.

Although the $\mathcal{O}(\alpha_s)$ perturbative corrections are not explicitly included in this analysis,  
the Wilson coefficients remain unaffected by temperature, as long as the temperature is well below the OPE separation scale (\(\sim 1\,\mathrm{GeV}\)) \cite{Hatsuda:1992bv}.
Thus, the temperature dependence of the charmonium spin structure arises predominantly from modifications of the gluon condensates $G_0$ and $G_2$. Consequently, the qualitative behavior observed in this work  is expected to remain valid even upon inclusion of these perturbative corrections.
For future studies, it will be essential to explicitly incorporate the spin decomposition of \(\mathcal{O}(\alpha_s)\) corrections and to explore additional medium effects beyond the dimension-4 gluon condensates. Such improvements will enable a more comprehensive and precise understanding of the charmonium spin structure in a hot medium.
%\newpage
\appendix
\section{Wilson coefficients of other operators}
For completeness, we provide the spin decomposed Wilson coefficients for the perturbative part and  dimension-4 scalar gluon operator $G_0$. The perturbative part can be written in terms of its imaginary part via the  dispersion relation,
\begin{align}
    \Pi^\text{pert}(Q^2)=\frac{1}{\pi}\int^\infty_{4m^2}ds\frac{\text{Im}\Pi^\text{pert}(s)}{s+Q^2}.
\end{align}
\hfill\break
\noindent $\bullet$ Vector channel: 
\begin{align}
    &\text{Im}\Pi^\text{pert}_{S_q}(s)=\frac{3m^2}{2\pi\sqrt{s(s-4m^2)}},\\
    &\text{Im}\Pi^\text{pert}_{L_q}(s)=\frac{(s-m^2)\sqrt{s(s-4m^2)}}{2\pi s^2},\\
    &C_{S_q}^{4,0}=\frac{-1}{12Q^4}\big\{2-y+12J_2-26J_3+12J_4\big\},\\
    &C_{L_q}^{4,0}=\frac{-1}{36Q^4}\big\{1+3y-(2+2y)J_1-5J_2+6J_3\big\},\\
    &C_{J_g}^{4,0}=\frac{1}{36Q^4}\big\{13-(2+2y)J_1-23J_2+12J_3\big\}.
\end{align}
\hfill\break
\noindent $\bullet$ Axial vector channel: 
\begin{align}
    &\text{Im}\Pi^\text{pert}_{S_q}(s)=\frac{3m^2\sqrt{s(s-4m^2)}}{2\pi s^2},\\
    &\text{Im}\Pi^\text{pert}_{L_q}(s)=\frac{(s-m^2)\sqrt{s(s-4m^2)}}{2\pi s^2},\\
    &C_{S_q}^{4,0}=\frac{-1}{12Q^4}\big\{6-y-12J_2+6J_3\big\},\\
    &C_{L_q}^{4,0}=\frac{-1}{36Q^4}\big\{1+3y+(6-2y)J_1-21J_2+14J_3\big\},\\
    &C_{J_g}^{4,0}=\frac{1}{36Q^4}\big\{1+(6-2y)J_1-3J_2-4J_3\big\}.
\end{align}

\section{Borel window and continuum threshold}\label{appendix2}
We also list the values of the Borel window and the threshold parameter employed in our spin decomposition analysis.
\hfill\break
\hfill\break
%\noindent $\bullet$ Vector channel: 
\begin{table}[H]
    \caption{Parameters for vector $(1^{--})$ state}
    \centering
    \begin{tabular}{ccc}
    \toprule
    $T/T_c$ & $\;\;\sqrt{s_0}$ [GeV]& \;\;$(M_{min},M_{max})$ [GeV]\\
    \midrule
    vacuum &3.57  & (1.08,2.35) \\
    1.00 & 3.47 &  (1.04,2.21)\\
    1.01 & 3.43 & (1.02,2.15) \\
    1.02 & 3.39 &  (0.99,2.07)\\
    1.03 & 3.34 & (0.97,1.98) \\
    1.04 & 3.29 & (0.95,1.90) \\
    \bottomrule
    \end{tabular}
    \label{tab:my_label}
\end{table}
%\noindent $\bullet$ Axial vector channel:
\begin{table}[H]
    \caption{Parameters for axial vector $(1^{++})$ state}
    \centering
    \begin{tabular}{ccc}
    \toprule
    $T/T_c$ & $\;\;\sqrt{s_0}$ [GeV]& \;\;$(M_{min},M_{max})$ [GeV]\\
    \midrule
    vacuum &  4.02& (1.39,2.33) \\
    1.00 & 3.87 & (1.32,2.14) \\
    1.01 & 3.79 & (1.28,2.04)  \\
    1.02 & 3.70 & (1.23,1.92) \\
    1.03 & 3.58 & (1.19,1.76) \\
    1.04 & 3.45 & (1.14,1.57) \\
    \bottomrule
    \end{tabular}
    \label{tab:my_label}
\end{table}
\begin{acknowledgments}
This work is supported by the WPI program “Sustainability with Knotted Chiral Meta Matter (WPI-SKCM$^2$)” at Hiroshima University.
\end{acknowledgments}

\bibliography{reference}

\end{document}